\documentclass[12pt,twocolumn, aps,nofootinbib,preprintnumbers,superscriptaddress,numberedappendix]{openjournal}
\usepackage{natbib}

\usepackage{newtxtext,newtxmath}

\usepackage[colorlinks=true, allcolors=blue]{hyperref}

\usepackage{graphicx}	% Including figure files
\usepackage{amsmath}	% Advanced maths commands
\newcommand{\orcidauthor}[3]{\author{\href{http://orcid.org/#1}{#2$^{#3}$}}}

\shorttitle{Lucky Image Technique on LOT}
\shortauthors{Hshieh \& Ngeow}
\begin{document}

%%%%%%%%%%%%%%%%%%% TITLE PAGE %%%%%%%%%%%%%%%%%%%

\title{Commissioning of a Commercial CMOS Camera for the Application of Lucky Image Technique on the Lulin One-meter Telescope}

\author{Yang-Peng Hsieh$^1$}
\orcidauthor{0000-0001-8771-7554}{Chow-Choong Ngeow}{1, *}

% Corresponding authors
\thanks{$^*$ Corresponding Author: \href{mailto:cngeow@astro.ncu.edu.tw}{cngeow@astro.ncu.edu.tw}.}

% Institutions
\affiliation{$^{1}$Graduate Institute of Astronomy, National Central University, 300 Jhongda Road, 32001 Jhongli, Taiwan}

%%%%%%%%%%%%%%%%%%%%%%%%%%%%%%%%%%%%

% Abstract of the paper
\begin{abstract}
Lucky image (LI) is a technique to achieve near diffraction-limit high-angular resolution images for meter-class optical telescopes. In this work, by observing the core of globular cluster M15, we demonstrated the LI technique can be applied to a 1-meter telescope, the Lulin One-meter Telescope (LOT), together with a commercial-grade CMOS camera. We have also developed a method to sort the quality of the LI frames by measuring the mean intensity per pixel on the selected reference stars. For a LI-reconstructed image based on the best 10\%-selected frames, we achieved a $1.7\times$ improvement on the full-width at half-maximum over the conventional long-exposure image. When cross-matched the detected sources on the LI-reconstructed image to the Gaia Data Release 3 catalog, we obtained a mean difference of $-0.04\pm0.09\arcsec$ and $-0.02\pm0.09\arcsec$ on the right ascension and declination, respectively, as well as reaching to a $5\sigma$ depth of $\sim 17.9$~mag in the Gaia $G$-band. 

% Max 250 words, one paragraph
\end{abstract}

\maketitle

\section{Introduction}\label{sec1}

Compared to adaptive optics, lucky image (LI) is a relatively inexpensive technique to achieve near diffraction-limited resolution for meter-class optical telescopes. The principles behind LI technique have been discussed elsewhere \citep[for examples, see][]{fried1978,baldwin2001,law2006,baldwin2008,hippler2009,smith2009,mackay2013,brandner2016}, and will not be repeated here. Simply speaking, by taking a series (typically in the order of $\sim 10^4$ to $\sim 10^5$) of very short exposure images, there is a chance (or ``luck'') such that a small fraction of these images would be least affected by atmospheric turbulence. Hence, by stacking these sharp images it is possible to produce an image which is close to the diffraction limit of the telescope.

In recent years, LI technique has been applied to various studies required high-resolution imaging. These studies ranged from binaries or multiple systems \citep{law2006b,law2008,ma2010,labadie2011,janson2012,rica2012,cc2017,janson2017,calissendorff2022}, host stars of exoplanets \citep{ginski2012,lb2012,bergfors2013,faedi2013,wollert2015,wollert2015b,evans2016,evans2018}, microlensing \citep{sajadian2016}, to globular clusters \citep{ds2012,skottfelt2013}. Other possible science cases enabled with LI technique have been discussed in \citet{mackay2018}. These LI studies used a variety of high-speed and low-noise CCD cameras and mounted on a number of 2 to 4 meter-class telescopes, such as the FastCam \citep[on various telescopes;][]{oscoz2008,rr2008}, the Two Colour Instrument \citep{skottfelt2015} on the Danish 1.54-m Telescope, the AstraLux \citep[Norte and Sur unit on 2.2-m Calar Alto Telescope and 3.6-m New Technology Telescope, respectively;][]{hormuth2008,hippler2009,janson2012}, the LuckyCam \citep{mackay2004,law2006} on the 2.56-m Nordic Optical Telescope (NOT), and the GravityCam \citep{mackay2018} on the 3.6-m New Technology Telescope. 

The above custom made high-speed CCD cameras are rather expensive. On the other hand, low-cost commercial CMOS cameras are getting popular among amateur astronomers. Therefore, it is desired to investigate the feasibility of applying commercial grade cooled-CMOS camera on meter-class telescope for LI-related studies. Hence, the main goal of this work is to test and commission such a commercial CMOS camera on the Lulin One-meter Telescope (LOT), and demonstrate this combination is capable to deliver a near diffraction-limited image. We selected ZWO ASI294MM-Pro (hereafter ASI294MM) CMOS camera, which can be cooled to $-35^\circ$ below ambient temperature. This CMOS camera can reach to a peak quantum efficiency of $\sim 90\%$, while maintaining a low noise level.

\begin{figure}
  \epsscale{1.0}
  \plotone{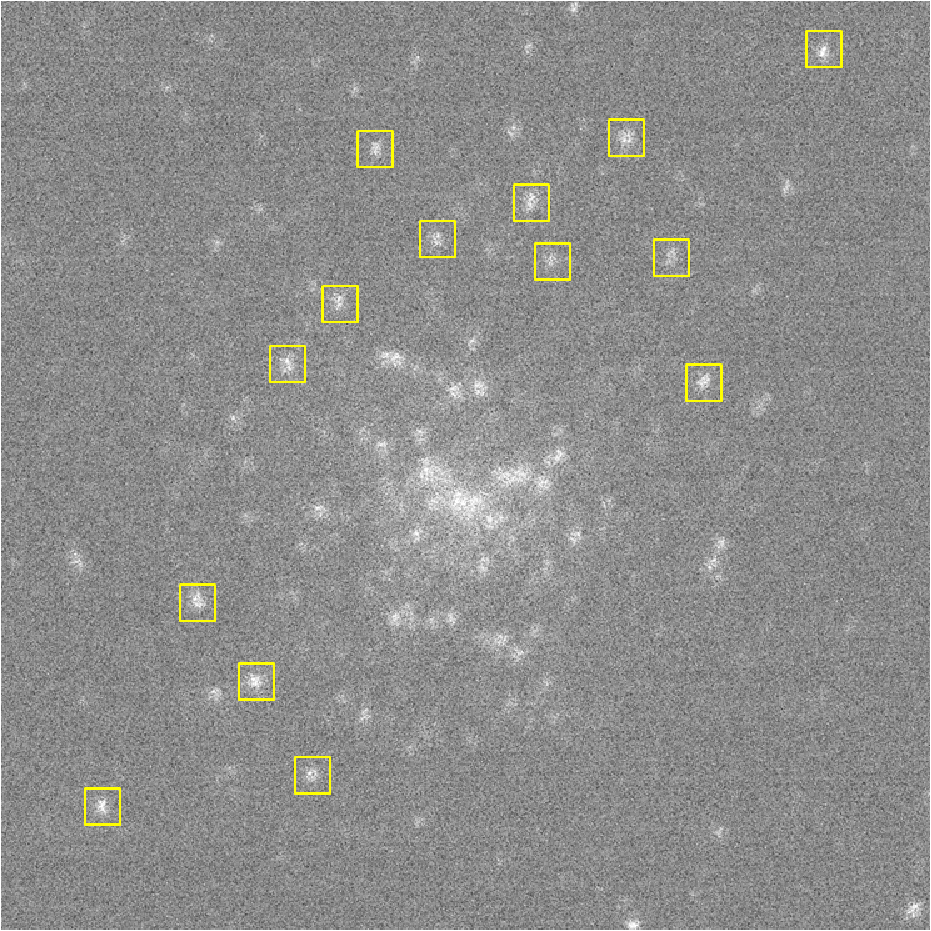}
  \caption{Locations of the 14 selected reference stars, marked in yellow squares, on a single 50~ms frame. The frame has a field-of-view (FOV) of $2.0\arcmin \times 2.0\arcmin$.}
  \label{fig_14ref}
\end{figure}

\section{LOT Observations and Image Reductions}\label{sec2}

\begin{figure*}
  \epsscale{1.3}
  \plotone{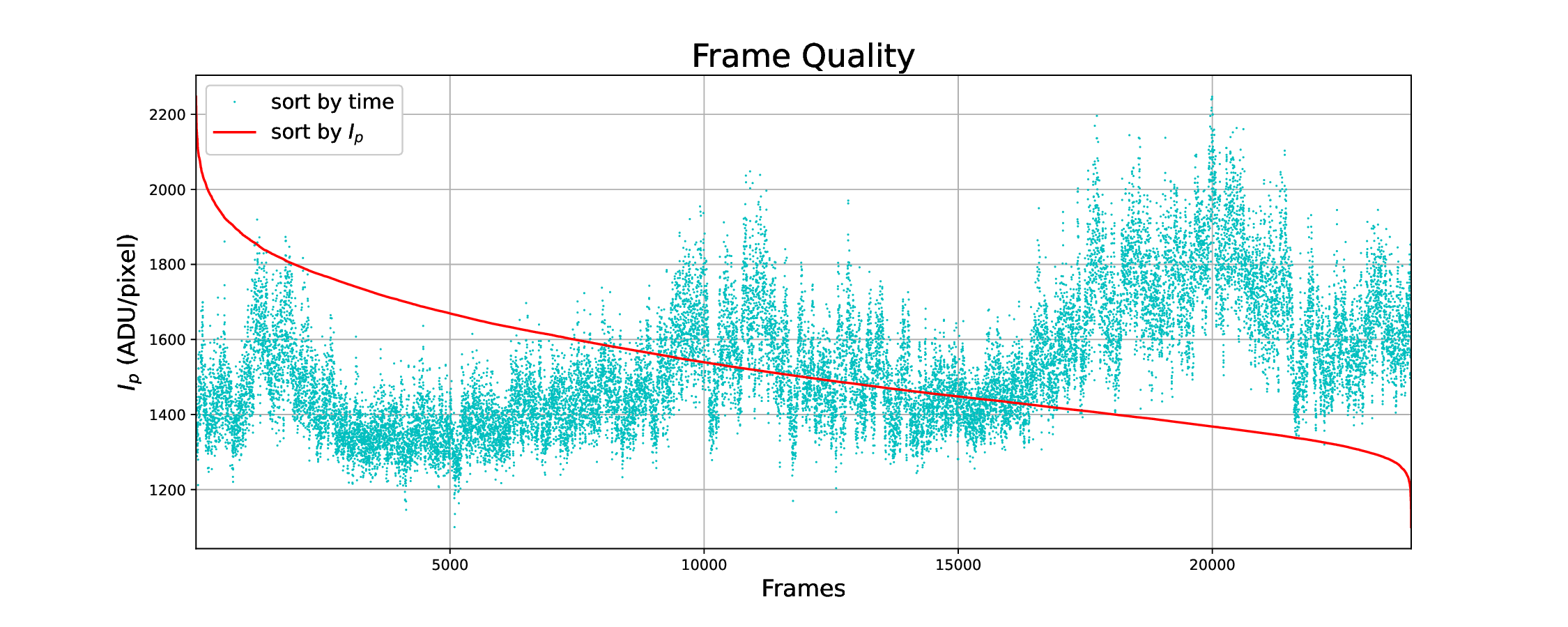}
  \caption{The pale-blue dots show the measured $I_p$ (see text for details) on each frames as a function of frame numbers (which also represent the time-sequence of the LI observations). The red curve represents the sorted $I_p$ values, which will be used to select the best quality frames.}
  \label{fig_Ip}
\end{figure*}

LOT is the largest research-grade telescope at the Lulin Observatory, located in central Taiwan with an elevation of 2862-m. LOT is a F/8 Cassegrain reflector, resulting a pixel scale of $0.12\arcsec/$pixel when equipped with the ASI294MM camera. This pixel scale is slightly smaller than the diffraction limit of $0.16\arcsec$ for LOT.

Our test observation was carried on 06 October 2021, by observing the core of globular cluster M15. The exposure time on a single frame was set to 50~milli-second (ms), i.e. 20~FPS (frame-per-second), with a total duration of 20 minutes. This resulted 24000 (science) frames with a data volume of $\sim 36GB$. To avoid crashing the data acquisition software, we divided the total duration into five segments. The entire LI observation was done via a commercial ASTRODON $r$-band filter.    

On the same night, we also took 4800 dark frames with the same 50~ms exposure time. These dark frames were median-combined to a master dark frame after applying a $5\sigma$ rejecting algorithm, and subsequently the master dark frame was used to subtract out the dark current in all of the science frames.

\section{Applying LI Technique}\label{sec3}

A key procedure in LI technique is to sort the acquired frames according to a given criterion or algorithm \citep[for example, the peak-pixel algorithm as presented in][]{aspin1997,smith2009,harpsoe2012}, such that the best quality frames can be stacked to produce a high-resolution image. In this work, we have selected 14 bright and relatively isolated stars as reference stars (see Figure \ref{fig_14ref}), and measured the total intensity within a $50\times 50$ pixels area centering on these reference stars. We defined $I_p$ (which has a unit of ADU/pixel) as the mean value of the measured intensity per pixel of these reference stars, and the best quality frame would have the largest $I_p$. In Figure \ref{fig_Ip}, the pale-blue dots represent the measured $I_p$ for each frames, reflecting the random-variation of instantaneous seeing during the observation. The sorted $I_p$ values are shown as the red curve in Figure \ref{fig_Ip}.

\begin{figure*}
  \epsscale{1.1}
  \plottwo{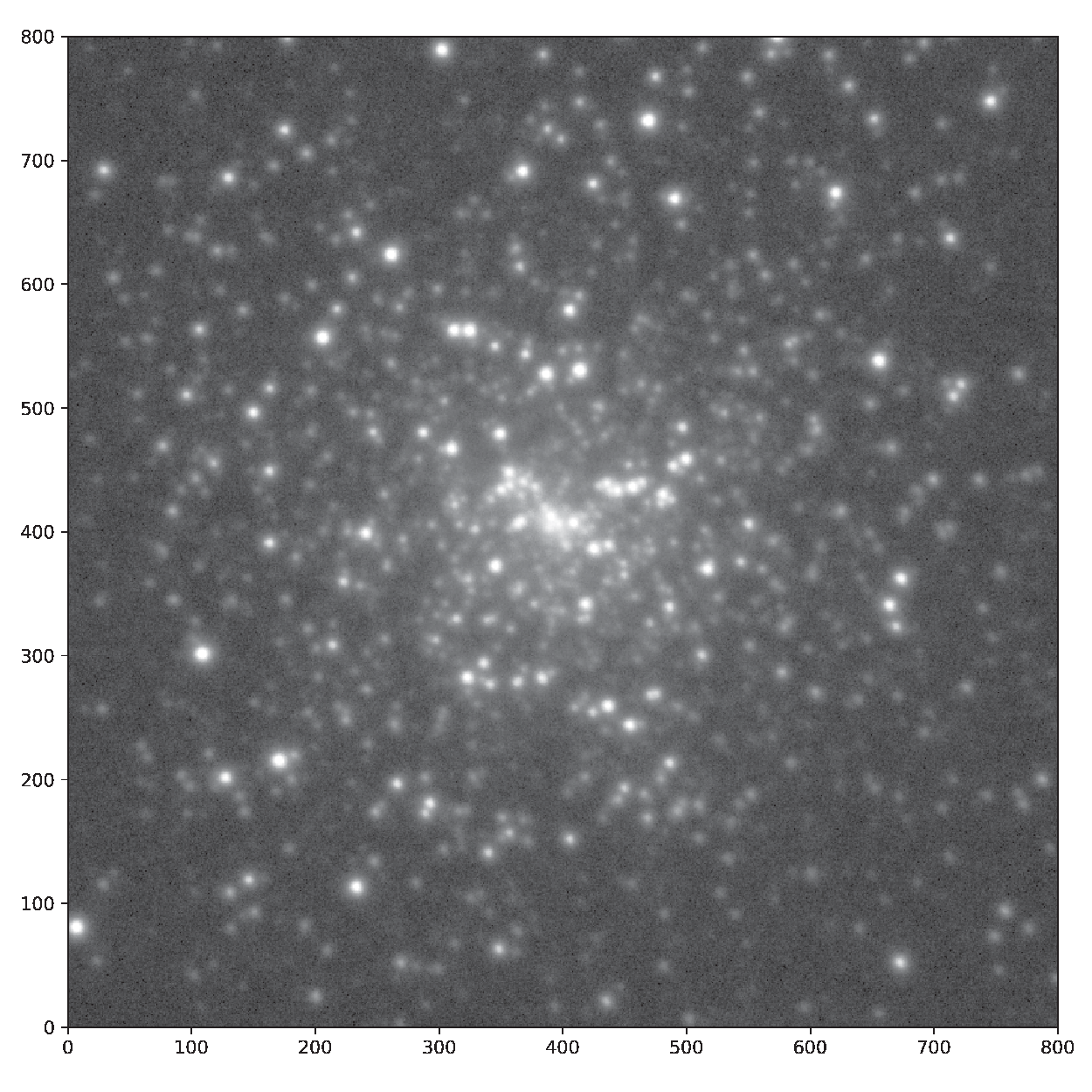}{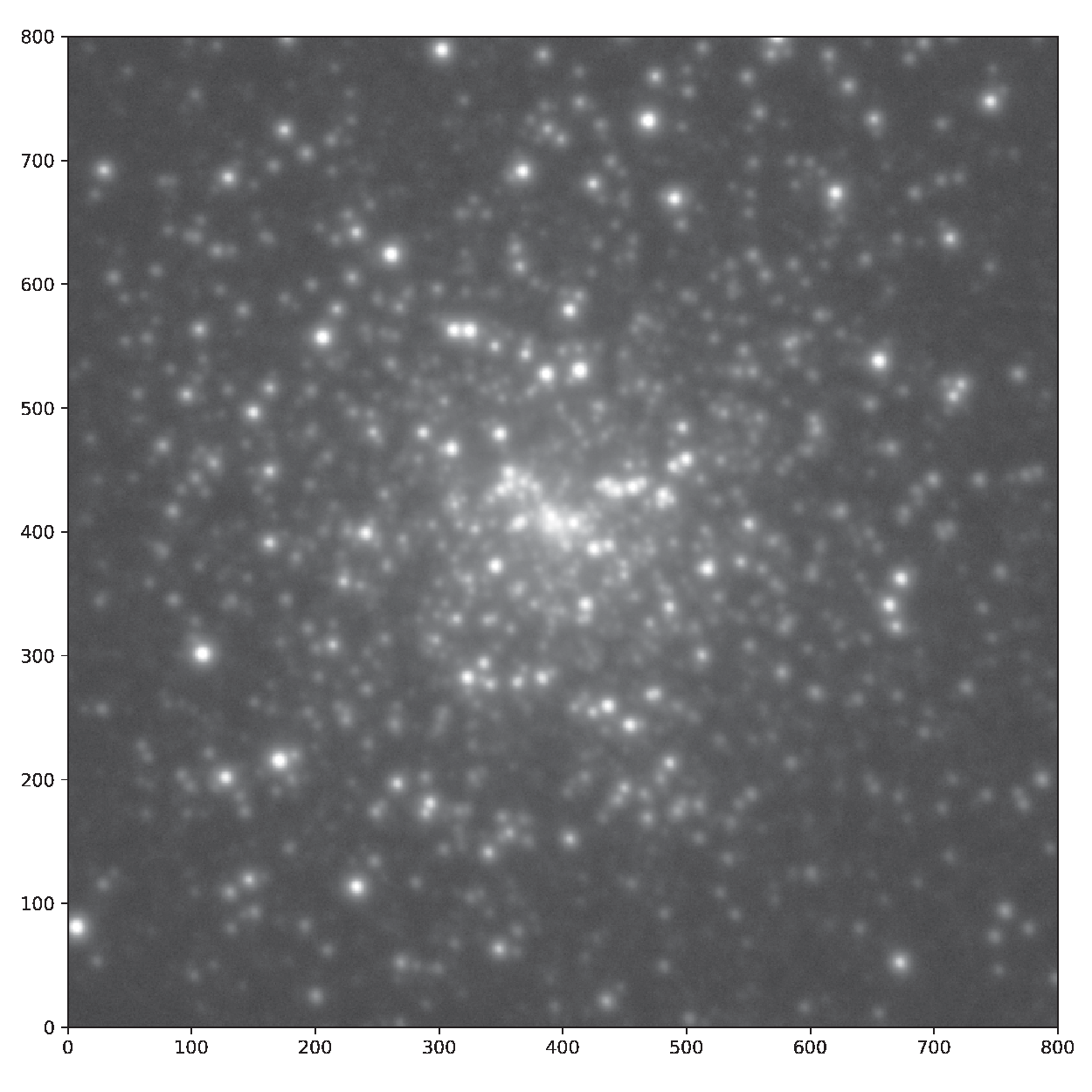}
  \plottwo{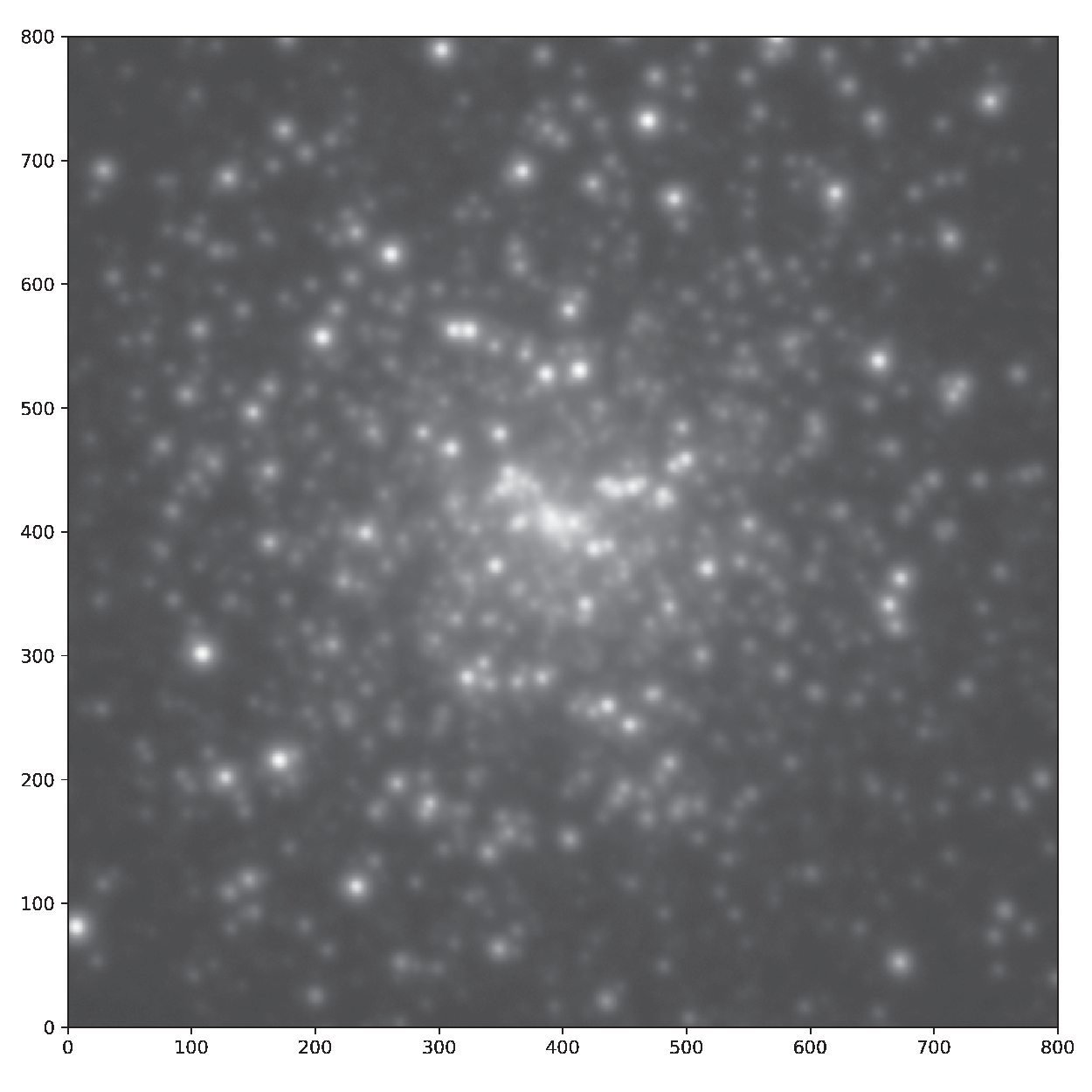}{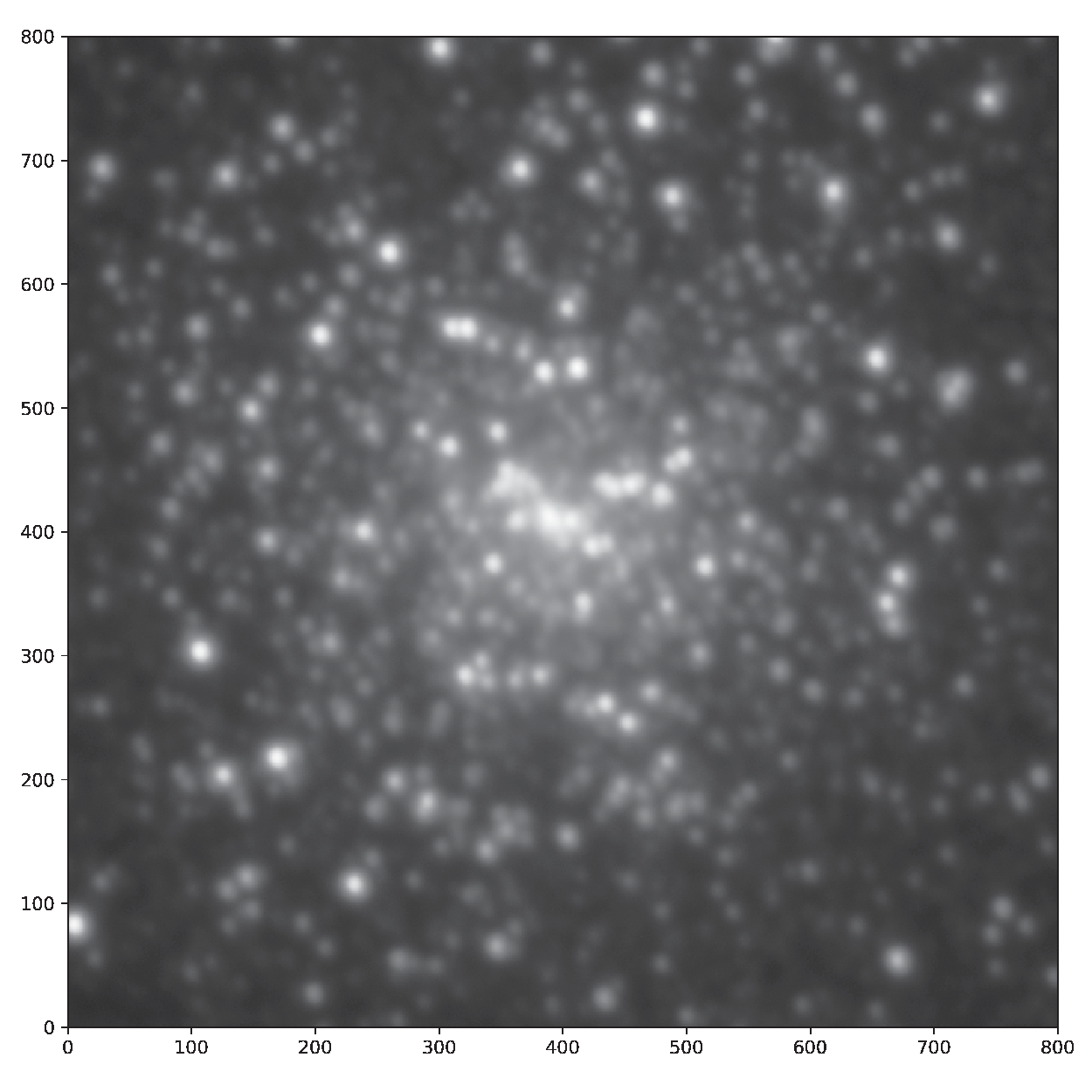}
  \caption{Examples of the stacked images based on various selection criterion in LI technique, including the selection of the best 1\% images (top-left panel), the best 10\% images (top-right panel), and full 100\% images (bottom-left panel), as compared to a 240-second long-exposure image displayed in the bottom-right panel.}% For illustration purpose, we picked the best 1\%, 10\%, and full 100\% selection of the frames.}
  \label{fig_li}
\end{figure*}

The next step is to define a reference point such that the selected best frames can be properly aligned. While measuring $I_p$ for the 14 reference stars in previous step, we have also measured the positions of the peak pixels (hereafter peak-positions) and the centroid of the reference stars from the best 2.5\% frames, and calculated the differences between these peak-positions and centroids. We further selected the top 0.5\% (or 120) frames that have the smallest mean differences, and took an average of the peak-positions for these 14 reference stars as the final reference point. Once we determined the reference point, we applied the classical shift-and-add method \citep[e.g.,][]{tubbs2002,harpsoe2012} to stack the best quality frames. To further improve the image reconstruction, we adopted a $4\times$~sub-pixel resampling algorithm and cropped the frames\footnote{This is to avoid crashing the computer while running the shift-and-add procedure.} into $1.6\arcmin \times 1.6\arcmin$ (as a result, two of the reference stars near the edges of the CCD were eliminated) while performing the shift-and-add procedure. 

\section{Analysis and Results}\label{sec4}

\begin{figure}
  \epsscale{1.15}
  \plotone{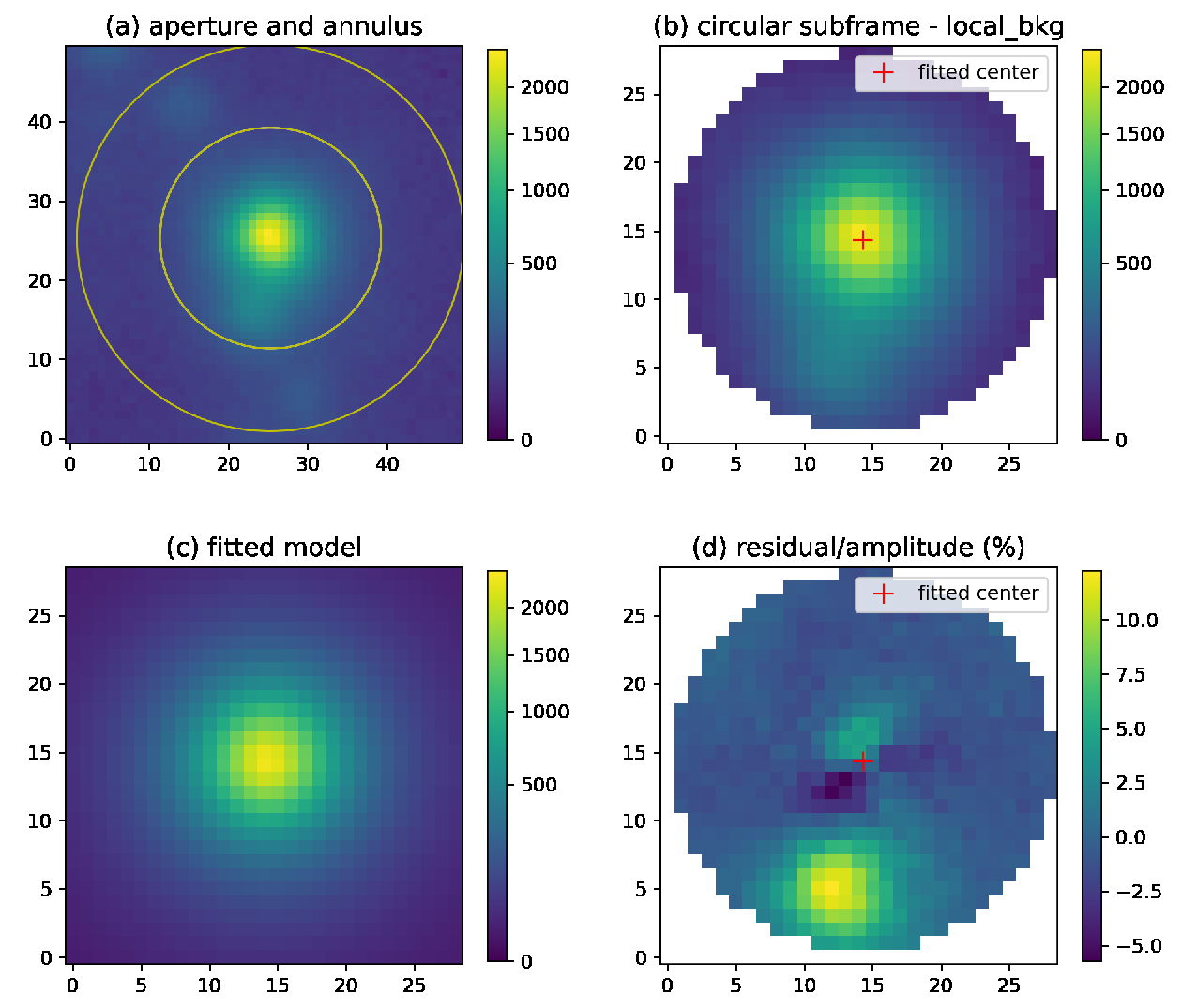}
  \caption{An example of reference star at which a faint neighboring star was revealed after subtracting the PSF model of the reference stars. {\bf Panel (a):} The aperture and the annulus for PSF fitting and local background estimation, respectively. {\bf Panel (b):} After subtracting the local background, we fit the Moffat profile, i.e. equation (1), to the reference star in the circular subframe. {\bf Panel (c):} The best-fit circular Moffat profile. {\bf Panel (d):} Residuals of the best-fit Moffat profile after scaling with the amplitude of the profile. Color-bars in (a) to (c) are in unit of ADU, while for (d) it is in percentage. For this reference star, a faint neighboring star can be clearly seen in (d) after subtracting the best-fit profile.}
  \label{fig_badrefstar}
\end{figure}

\begin{figure}
  \epsscale{1.15}
  \plotone{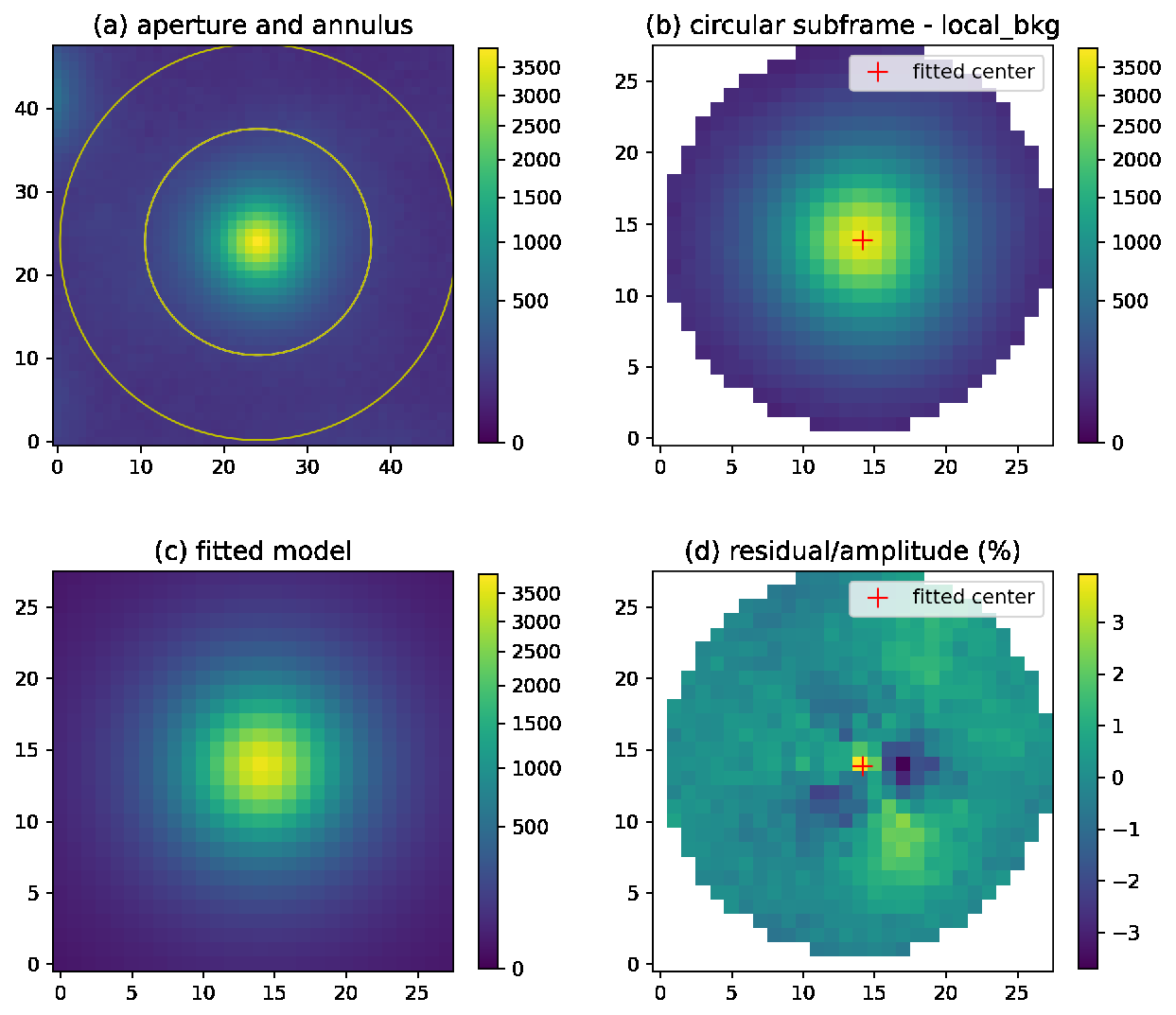}
  \caption{Same as Figure \ref{fig_badrefstar}, but for a reference star without a faint neighboring star. This star is selected as the standard star as described in the text.}
  \label{fig_goodrefstar}
\end{figure}

Figure \ref{fig_li} presents the stacked images by selecting the best 1\%-, 10\%-, and 100\%-selected frames (hereafter LI-reconstructed images), as well as a 240~second long exposure image (taken on the same night, with the same camera and filter on LOT). Improvements on the image quality (such as the ``sharpness'' of the point sources) via LI technique can be clearly seen when compared to the conventional long-exposure image. 

\subsection{FWHM Improvement}

We first evaluate the improvement of the full-width at half-maximum (FWHM) of the stellar PSF profile in the LI-reconstructed images against the conventional long-exposure image, by using the same 12 reference stars selected in previous section. The stellar PSF profile was fitted using a circular Moffat profile \citep{moffat1969,bendinelli1988,trujillo2001}:

\begin{eqnarray}
  M(r) & = & \frac{\gamma -1 }{\pi \alpha^2} \left[ 1 + \left(\frac{r}{\alpha}\right)^2 \right]^{-\gamma},
\end{eqnarray}

\noindent which depends on two fitted-parameters $\alpha$ and $\gamma$. For such a profile, it can be shown that $FWHM=2\alpha \sqrt{2^{1/\gamma}-1}$.

\begin{figure}
  \epsscale{1.1}
  \plotone{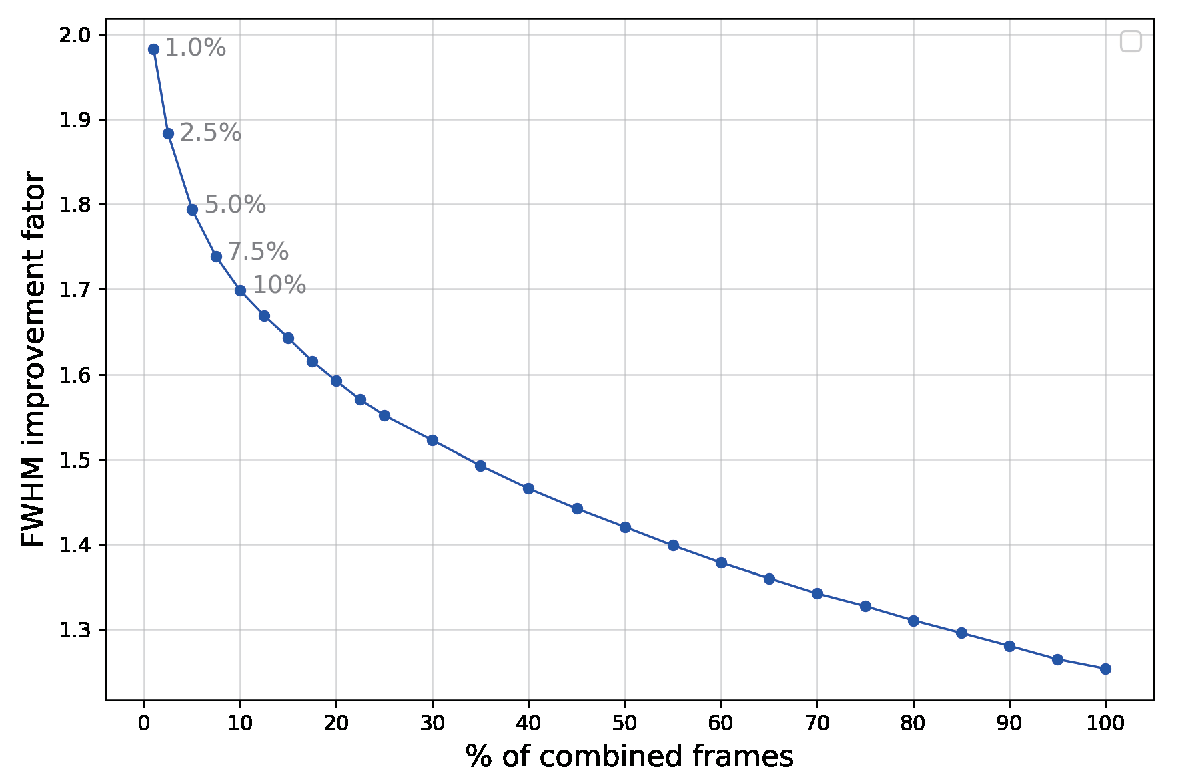}
  \caption{The FWHM improvement factor as a function of the percentage to select the best quality frames.}
  \label{fig_fwhm}
\end{figure}

Given the crowded nature of globular clusters, especially at the core region, there is a non-negligible possibility that the reference stars could have faint and yet unresolved neighbors contributing to their PSF profiles. Therefore, when fitting the Moffat profile to the reference stars, we adopted a circular region instead of a square region \citep[such as the one implemented in the {\tt photutils} package,][]{bradley2022} to minimize contamination from the faint and unresolved neighboring stars. The aperture of such circular region was set to $2\times FWHM$, with an annulus of the size of $2$-$3.5\times FWHM$ for local background estimation. We provided an initial guess of $FWHM$, and our {\tt python} scrip would improve the $FWHM$ in a few iterations until the solutions converged. Figure \ref{fig_badrefstar} and \ref{fig_goodrefstar} illustrate two examples of the PSF profile fitting to the reference stars.

For five of the reference stars, a nearby faint companion can be seen after subtracting the best-fit Moffat profile, Figure \ref{fig_badrefstar} shows one of such example. In fact, the presence of a faint and unresolved companion can also be inferred from the fitted Moffat parameters $\alpha$ and $\gamma$, especially for $\gamma$, as their values shows a large deviation ($1.69 \leq \alpha \leq 2.22$ and $4.84 \leq \gamma \leq 5.83$). Another seven reference stars do not have a nearby unresolved companion, demonstrated in Figure \ref{fig_goodrefstar} as an example. These seven reference stars share similar $\alpha$ and $\gamma$ values, with an average of $\langle \alpha \rangle = 2.13\pm0.10$ and $\langle \gamma \rangle = 5.57\pm0.27$, respectively, and will be adopted as the standard stars for measuring the FWHM of the LI-reconstructed images.

We defined a FWHM improvement factor to be the ratio of the nominal seeing and the FWHM measured from the LI-reconstructed images, where the nominal seeing measured from the conventional long-exposure image is $1.38\arcsec$. A larger FWHM improvement factor implies the FWHM is better than the nominal seeing. Figure \ref{fig_fwhm} presents the FWHM improvement factor as a function of the percentage when selecting the best frames. We achieved a maximum FWHM improvement factor of 1.98 for the best 1\%-selected frames, and a minimum of 1.25 when all frames were selected. Note that the FWHM for the 100\%-selected frames is better than the conventional long-exposure image, because the frames have been shifted and added using the procedures mentioned in Section \ref{sec3}.

\begin{figure}
  \epsscale{1.1}
  \plotone{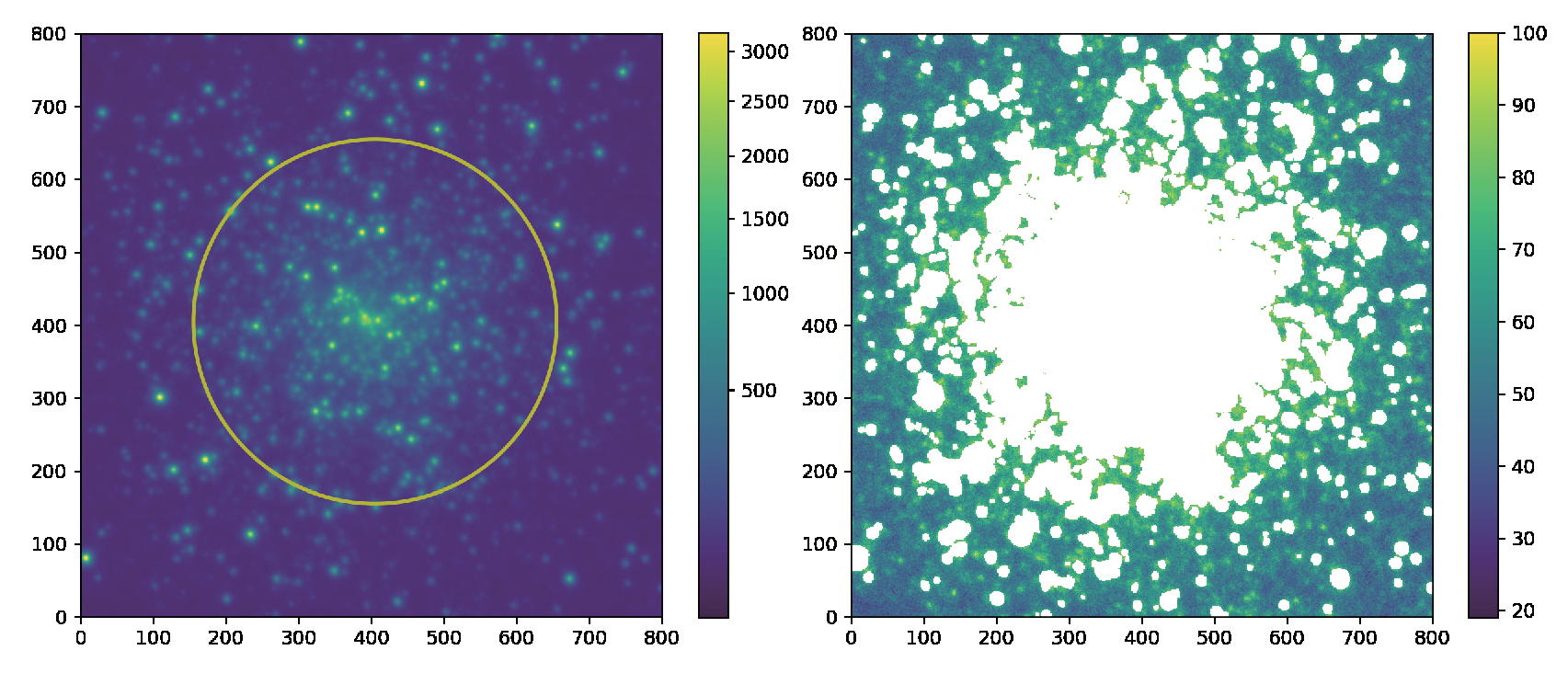}
  \caption{Sky region for global background estimation, where the inner region (marked with the yellow circle on the left panel) and the stellar sources were masked out.}
  \label{fig_bkg}
\end{figure}

\subsection{Sources Detection}

We applied the {\tt DAOStarFinder} code, implemented in the {\tt photutils} package, which is based on the classical {\tt DAOFIND} algorithm \citep{stetson1987}, for detecting sources in the LI-reconstructed images. A crucial step for sources detection is background estimation. Even though the field-of-view (FOV) of our images is small ($1.6\arcmin \times 1.6\arcmin$), the global background is not flat but brighter at the center of the image. This is due to the blended faint stars at the core of M15 from the increasing of stellar density, hence the background can vary from $\sim50$~ADU to $\sim 400$~ADU at the edge and the center of the image, respectively.

Using the LI-reconstructed image from the best 10\%-selected frames as an example (and throughout the paper), we first measured the median and standard deviation of the background in the region outside the yellow circle shown in the left panel of Figure \ref{fig_bkg} as initial guess. We then applied the {\tt make\_source\_mask} code from the {\tt photutils} package to mask out sources and the central region, as displayed in the right panel of Figure \ref{fig_bkg}. The median of the final background level and the associated standard deviation ($\sigma_{BG}$) was determined from the unmasked region. We then ran the {\tt DAOStarFinder} with a $3\sigma_{BG}$ detection threshold. Note that {\tt DAOStarFinder} employed a Gaussian kernel to detect sources, hence we determined the Gaussian FWHM (and not the Moffat FWHM) using the seven standard stars mention previously. On this best 10\%-selected and LI-reconstructed image, we detected 737 sources. 

\begin{figure}
  \epsscale{1.1}
  \plotone{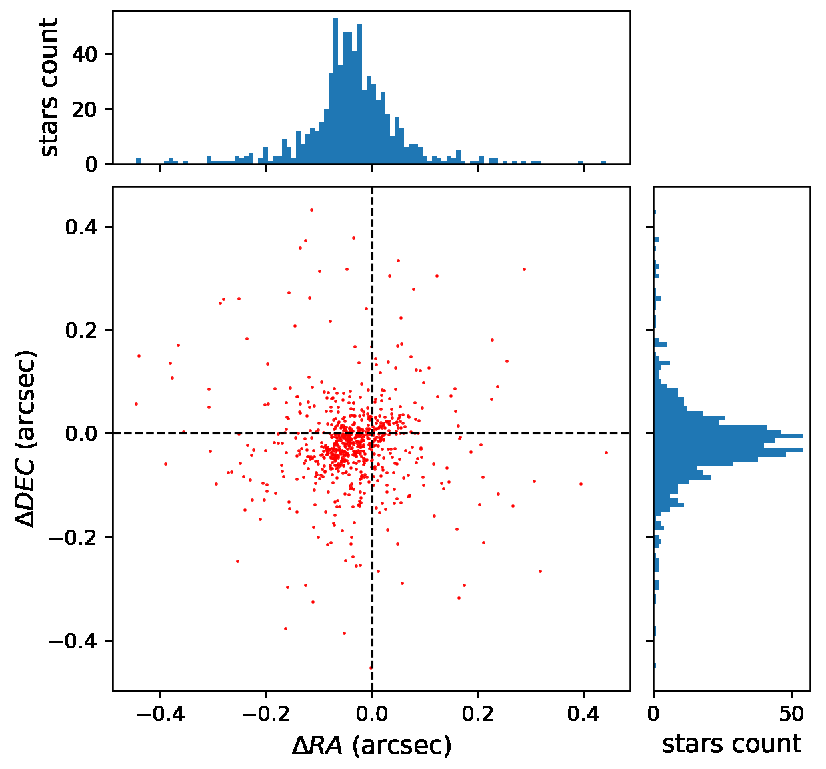}
  \caption{Deviations of the right ascension ($RA$) and declination ($DEC$) for the sources matched to the Gaia DR3 catalog.}
  \label{fig_radec}
\end{figure}

\begin{figure}
  \epsscale{1.1}
  \plotone{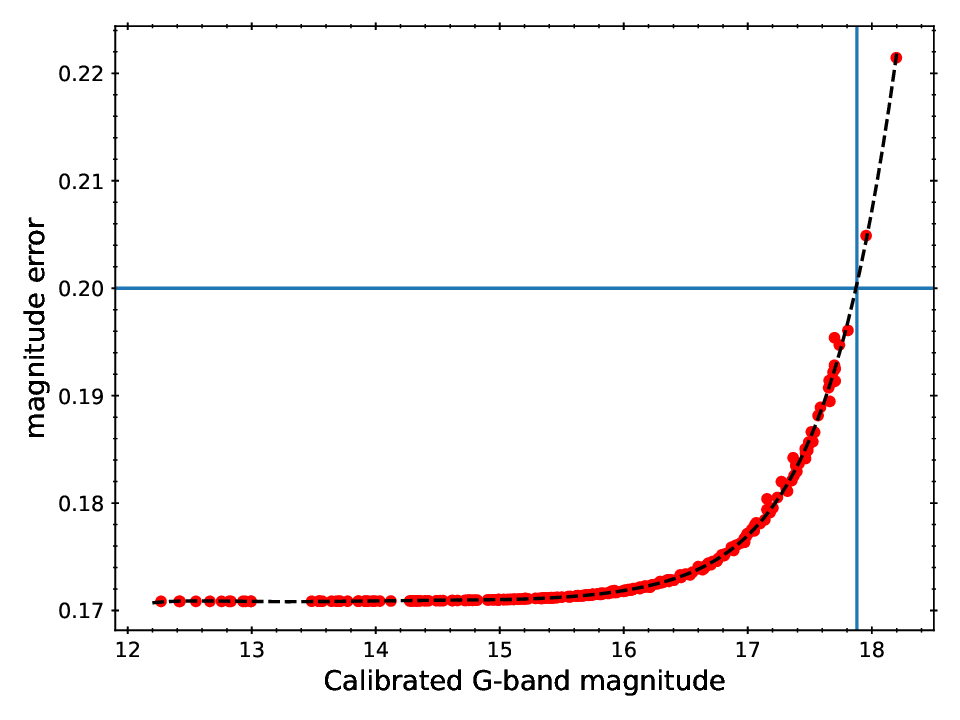}
  \caption{Photometric errors as a function of calibrated Gaia $G$-band magnitudes for the detected sources in the best 10\%-selected and LI-reconstructed image. The black dashed curve represents a polynomial fit to the data, which was used to determine the $5\sigma$ limiting magnitude (illustrated with the blue lines).}
  \label{fig_limitmag}
\end{figure}

\subsection{Astrometry and Photometry}

We used the {\tt solve-field} code, implemented in the {\tt astrometry.net} package \citep{lang2010}, to solve the astrometry for the sources detected in the best 10\%-selected and LI-reconstructed image. We adopted the Gaia Data Release 3 \citep[DR3,][]{gaia2016,gaia2022} as the astrometric reference catalog.\footnote{We custom built the index file from the Gaia DR3 catalog that is required for running the {\tt solve-field} code.} For the 737 detected sources, 717 of them matched to the entries in the Gaia DR3 reference catalog, and we achieved the following sub-arcsecond accuracy in right ascension ($RA$) and declination ($DEC$): $\langle \Delta RA \rangle = -0.04\pm0.09\arcsec$ and $\langle \Delta DEC\rangle = -0.02\pm0.09\arcsec$ (see Figure \ref{fig_radec}).

From the matched Gaia sources, we further selected $\sim100$ ``good-quality'' stars for photometric calibration. These ``good-quality'' stars are those not saturated in the image, located neither near the edge of the image nor near the bright inner region, and do not have relatively bright and nearby neighbors. An iterative sigma-clipping linear least-square fitting procedure was used to solve for the zero-point ($ZP$) and the color-term ($C$) in the following calibration equation: $G-r_{\mathrm{instr}} = ZP + C (B_p-R_p)$, where $G$ are the Gaia $G$-band magnitudes, $(B_p-R_p)$ are the colors in Gaia's $B_p$ and $R_p$ filters, and $r_{\mathrm{instr}}$ represents the PSF instrumental magnitudes for these ``good-quality'' stars. Once we solved the $ZP$ and the $C$ term, we applied the calibration equation to the detected sources in the image. Figure \ref{fig_limitmag} shows the errors on the magnitude as a function of the calibrated $G$-band magnitudes, and we determined the $5\sigma$ limiting magnitude in the $G$-band to be $17.88$~mag.

\section{Discussion and Conclusions} \label{sec5}

\begin{figure}
  \epsscale{1.1}
  \plotone{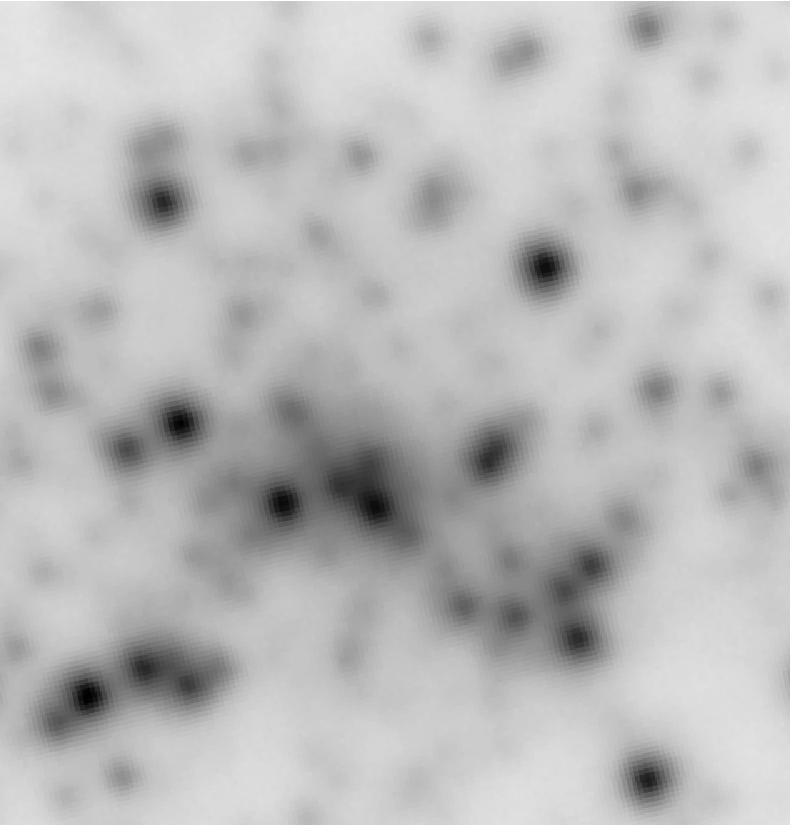}%{zoom.eps}
  \caption{Zoomed in of the top-right panel in Figure \ref{fig_li} for the comparison with Figure 1 shown in \citet[][their right panel]{law2006} and \citet[][their left panel]{ds2012}.}
  \label{fig_zoom}
\end{figure}

\begin{deluxetable}{lccc}
  %\movetableright=-1in
  \tabletypesize{\scriptsize}
  \tablecaption{Comparison of our work with \citet[][L2006]{law2006} and \citet[][DS2012]{ds2012}.\label{tab1}}
  \tablewidth{0pt}
  \tablehead{
    \colhead{} &
    \colhead{This Work} &
    \colhead{L2006} &
    \colhead{DS2012} 
  }
  \startdata
  Telescope  & 1-m LOT  & 2.56-m NOT & 2.56-m NOT \\
  Camera     & ASI294MM & LuckyCam & FastCam \\
  Diffraction limit & $0.16\arcsec$ & $0.08\arcsec$ & $0.08\arcsec$ \\
  Pixel scale& $0.12\arcsec/$pixel & $0.04\arcsec/$pixel & $0.03\arcsec/$pixel \\
  FOV        & $1.6\arcmin \times 1.6\arcmin$ & $22.0\arcsec \times 20.5\arcsec$ & $13.2\arcsec \times 13.2\arcsec$ \\
  Limiting magnitudes & $G\sim17.9$~mag & $\cdots$ & $I\sim19.5$~mag \\
  Single frame exposure & 50~ms & 83.3~ms & 30~ms \\
  Total duration & 20~min & $\cdots$  & 2~hr 43~min \\
  Nominal seeing & $1.38\arcsec$ & $\sim 0.63\arcsec$ & $\sim 0.65\arcsec$ 
  \enddata
\end{deluxetable}

In this work, we successfully commission and test the application of LI technique on LOT using a commercial CMOS camera. We demonstrated that such a combination can achieve $1.25\times$ to $\sim2\times$ improvement on the FWHM, reaching sub-arcsecond accuracy on the astrometry, and reaching a $5\sigma$ depth of G=17.88~mag, by using the $I_P$ as a measure of the frames quality. We notice that our LI observation on the core of M15 is similar to the work of \citet{law2006} and \citet{ds2012} (see Table \ref{tab1} for comparisons of some features between these studies). When zoomed in to our LI-reconstructed image (see Figure \ref{fig_zoom}), we recovered almost all of the bright sources. Certainly, the fainter sources in our LI-reconstructed image were either blended or undetected due to a smaller aperture of LOT as compared to NOT. Hence, a commercial-grade CMOS camera equipped on the increasing number of 1-m class telescopes \citep[for examples,][]{holtzman2010,hu2014,bai2020}, or other similar aperture telescopes, is capable for applying the LI technique to study various science cases as mentioned in the Introduction. 

%%%%%%%%%%%%%%%%%%%%%%%%%%%%%%%%%%%%%%%%%%%%%%%%%%
\section*{Acknowledgements} \label{sec:acknowledgements}
%%%%%%%%%%%%%%%%%%%%%%%%%%%%%%%%%%%%%%%%%%%%%%%%%%

%We thank the useful discussions and comments from an anonymous referee to improve the manuscript.
This publication has made use of data collected at Lulin Observatory, partly supported by NSTC grant 109-2112-M-008-001. This research has made use of the SIMBAD database and the VizieR catalogue access tool, operated at CDS, Strasbourg, France. This research made use of Astropy,\footnote{\url{http://www.astropy.org}} a community-developed core Python package for Astronomy \citep{astropy2013, astropy2018, astropy2022}. This research made use of Photutils, an Astropy package for detection and photometry of astronomical sources \citep{bradley2022}. This work has made use of data from the European Space Agency (ESA) mission {\it Gaia} (\url{https://www.cosmos.esa.int/gaia}), processed by the {\it Gaia} Data Processing and Analysis Consortium (DPAC, \url{https://www.cosmos.esa.int/web/gaia/dpac/consortium}). Funding for the DPAC has been provided by national institutions, in particular the institutions participating in the {\it Gaia} Multilateral Agreement.

%%%%%%%%%%%%%%%%%%%%%%%%%%%%%%%%%%%%%%%%%%%%%%%%%%

% Don't change these lines
\end{document}